\documentclass[aps,pra,reprint,twocolumn,amsfonts,amssymb,amsmath,floatfix,tightenlines]{revtex4-1}
\usepackage{graphicx}
\usepackage{pdfpages}
\usepackage{color}
\usepackage[colorlinks=true,citecolor=blue,urlcolor=blue]{hyperref}

\numberwithin{equation}{section}

\newcommand*\conj[1]{\bar{#1}}
\def\conj{\bar}
\def\corr{\mathrm{corr}}
\def\mb{\mathbf}\def\fl{\mathrm{fl}}\def\eff{\mathrm{eff}}
\def\p{\partial}\def\pg{\mathrm{pg}}\def\sc{\mathrm{sc}}
\def\BCS{\mathrm{BCS}}\def\mf{\mathrm{mf}}
\def\Tr{\mathrm{Tr}}\def\tr{\mathrm{tr}}

\begin{document}

\title{Gauge invariant theories of linear response for 
strongly correlated superconductors}

\author{Rufus Boyack}
\author{Brandon M. Anderson}
\author{Chien-Te Wu}
\author{K. Levin}
\affiliation{James Franck Institute, University of Chicago, Chicago, Illinois 60637, USA}

\date{\today}

\begin{abstract}
We present a general diagrammatic theory
for determining consistent electromagnetic response functions in strongly
correlated fermionic superfluids. 
The general treatment of correlations beyond BCS theory requires a new theoretical formalism
not contained in the current literature.
Among concrete examples are a rather extensive class of theoretical models
which incorporate BCS-BEC crossover as applied to
the ultra cold Fermi gases, along with theories
specifically associated with the high-$T_c$ cuprates. 
The challenge is to maintain gauge invariance, while simultaneously incorporating
additional self-energy terms arising from strong correlation effects.
Central to our approach is the application of
the Ward-Takahashi identity, which introduces collective
mode contributions in the response functions and guarantees
that the $f$-sum rule is satisfied. 
We outline a powerful and very general method to determine these collective modes
in a manner compatible with gauge invariance. 
Finally, as an alternative approach, we contrast with the path integral formalism. 
Here, the calculation of gauge invariant response appears more straightforward.
However, the collective modes introduced are essentially those of strict BCS theory,
with no modification from correlation effects.
Since the path integral scheme simultaneously addresses
electrodynamics and thermodynamics, 
we emphasize that it should be subjected to a
consistency test beyond gauge invariance, 
namely that of the compressibility sum-rule.
We show how this sum-rule  fails in the conventional path integral approach.
\end{abstract}

\maketitle

\section{Introduction}
There has been a recent focus in the literature
on strongly correlated superconductors and superfluids.
This interest has arisen in two different contexts,
via ultra cold atomic Fermi gases \cite{Taylor, Randeria1}
and via high-$T_c$ superconductors \cite{YRZ, Lee, Norman,ourreview}.
A major challenge in studying these two different systems
is to arrive at correct expressions for
the electromagnetic (EM) properties, such as the superfluid density and
the density-density correlation function, 
which characterize superconductors and superfluids.

In strict BCS theory there are two different conventional techniques
for addressing electromagnetic response while ensuring gauge invariance: 
the path integral \cite{Altland, Yakovenko, Kitagawa} and the Ward-Takahashi identity \cite{Nambu}.
The first of these methods depends on the derivation of a generating functional
while the second depends on the form of the diagrammatic self-energy.
This body of work has enabled a complete understanding of
the gauge invariant electromagnetic response at the BCS level.
It does not, however, answer the important questions about
how to incorporate stronger correlation effects. 

Studies of high-$T_c$ superconductors, which necessarily
require a beyond-BCS formalism, are better suited to the
Ward-Takahashi based approach.
These studies focus on different models 
for the self-energy associated with a
normal state that includes pairing, known as the pseudogap phase \cite{YRZ,Lee,Norman,ourreview}.
This correlation contribution to the
self-energy has been extensively characterized \cite{arpesstanford} above the
transition temperature $T_{c}$. 
In the superfluid phase, presumably
one adds to this normal state self-energy \cite{ourreview,YRZ}
an additional BCS self energy contribution.
The challenge in studying strongly correlated superfluids, however,
is ensuring gauge invariance. This means 
that the self-consistent collective modes,
compatible with gauge invariance, must be properly included.
For an arbitrary strongly-correlated self-energy, beyond the BCS-level theory, 
there is no general diagrammatic procedure to ensure both of these conditions.

In this paper we show that the self-energy and the gap equation
provide all the ingredients required to unambiguously establish the exact electrodynamic response
at all temperatures.
Our main goals are:

\textit{(i) To show how
to arrive at the exact gauge invariant electromagnetic response
of strongly correlated superfluids. This is based on a fairly
general form of the self-energy and on the Ward-Takahashi identity.} 

\textit{(ii) To provide a powerful method for obtaining
the collective modes in a gauge invariant manner for strongly correlated superfluids.
This is based on the form of the gap equation, and the vertex derived above in (i).}

The electrodynamics of superconductors is also widely addressed via the 
path integral approach \cite{Altland, Yakovenko, Kitagawa} which 
requires the introduction of Gaussian level (beyond saddle point) fluctuations.
Incorporating gauge invariance is relatively straightforward, 
which is in large part due to the fact that the collective modes
that enter at this level and beyond are those of strict BCS theory \cite{ourcompanion}.
We shall revisit this conventional calculation of response functions at the strict BCS level, 
while simultaneously considering thermodynamics.
We find there is a serious shortcoming that has not previously been identified in the literature. 
This arises from an inconsistency between
electrodynamics and thermodynamics, which is 
manifested as a failure of the compressibility sum-rule.

Our emphasis here is not on a critique of previous work since, quite generally,
in the literature the focus has been on either the thermodynamics \cite{Randeria1}
or the electrodynamics \cite{Altland, Yakovenko, Kitagawa},
but not on both simultaneously.
Nevertheless, the violation of the compressibility sum-rule
is a serious shortcoming.
The source of this sum-rule violation
comes from the fact that the BCS level electrodynamics are derived 
by incorporating beyond BCS Gaussian fluctuations.
This would seem to require that we also include Gaussian
fluctuations in the number equation. 
However, this in fact leads to the failure of the compressibility sum-rule.
A detailed discussion of how to implement consistency between
electrodynamics and thermodynamics will be presented elsewhere \cite{ourcompanion}.

It is crucial when studying transport phenomena to ensure that all
conservation laws, such as energy, momentum, and charge, are satisfied \cite{KadanoffBaym, Baym}.
In particular, ensuring gauge invariance, and thus charge conservation, in a superconductor has long been 
a problem of great importance \cite{Anderson, Kadanoff, Nambu, Kulik, Hao1}.
The key insight in the challenge of preserving gauge invariance,
even in the presence of a Meissner effect, was the necessity 
of long wavelength collective excitations \cite{Anderson, Rickayzen}. 
Following this initial insight, a more diagrammatic approach, built around
the establishment of gauge invariance
in quantum electrodynamics, was developed by Nambu \cite{Nambu}. 
Nambu's method of establishing a gauge invariant electromagnetic response
was to set up a gauge invariant vertex at the same level of approximation
as the self-energy. He then showed that this leads to a full vertex
that satisfies the Ward-Takahashi identity (WTI),
a condition equivalent to gauge invariance \cite{Ryder}.

A modern understanding of the role of gauge invariance in a 
superconductor is best understood from this field theoretic point of view:
collective modes are excitations which restore gauge invariance. 
In the language of quantum field theory they can be interpreted as the Nambu-Goldstone bosons 
arising from spontaneous symmetry breaking in the condensed phase.
Strictly speaking, in a superconductor or superfluid local gauge invariance is never broken \cite{Greiter}. 
Quite generally, the impossibility of breaking local gauge invariance without explicit 
gauge fixing, at least for abelian gauge fields, was proved early on by Elitzur \cite{Elitzur}.
Rather, due to the presence of a condensate, global phase invariance is spontaneously broken. 
In the case of a neutral order parameter
the excitation spectrum contains a gapless mode, which corresponds to 
the collectives modes discussed throughout this paper. 
For a charged order parameter the Goldstone modes couple 
to the longitudinal degrees of freedom of the gauge field, 
and are gapped out.

In going beyond the BCS theory of superconductivity it 
is essential that gauge invariance is maintained in any approximation scheme.
Above the transition temperature, in Refs. \cite{Peter1, Rufus1}
the WTI was implemented for a number of different exotic normal phases,
which led to a consistent framework for computing all vertex corrections. 
The challenge in the present paper is then to extend this body of work
and formulate a gauge invariant theory below the transition temperature. 
In this context, Ref. \cite{Hao2} used the WTI to formulate a gauge 
invariant response for a specific BCS-BEC approximation valid at all temperatures.
This theory accounted for non-condensed fermionic pairs by adding a 
$t$-matrix self-energy to the standard BCS self-energy.
Inspired by this work, in this paper we will use the WTI to study a broader class of theories,
addressed in the context of high $T_c$ superconductors and atomic
Fermi superfluids, which are based on an extension of a BCS based self-energy. 
Within these approaches we go beyond the pioneering work of Nambu and show 
by extending the method of Ref. \cite{Hao2}, that both the full vertex and the collective modes
can be explicitly derived for a very general class of strongly correlated superfluids.
In particular we derive closed form expressions for the response functions.
Theories which belong to this general class include the work of Refs. \cite{Peter1, Rufus1}
along with additional theories such as that proposed in Ref. \cite{YRZ},
Ref. \cite{Lee} and Refs. \cite{StrinatiPRL, StrinatiPRB, StrinatiPRB2}.

\section{Correlation Effects Beyond BCS theory: Ward-Takahashi Identity}
\label{sec:Ward}
\subsection{Kubo formulae}
The goal of this section of the paper is to address correlations which go beyond
the mean-field BCS theory and, making use of Kubo formulae,
arrive at properly gauge invariant linear response functions. 
We begin by summarizing the Kubo formalism
for a many-body theory of interacting fermions.
In what follows we shall primarily be concerned with neutral superfluids. 
Incorporating Coulomb effects can be done through the random phase approximation (RPA) formalism \cite{Yakovenko},
once the exact response functions are obtained for the neutral system. 

In the presence of a weak, externally applied EM field, with
four-vector potential $A^{\mu}=(\phi, \mathbf{A})$, the
four-current density $J^{\mu}=(\rho, \mathbf{J})$ is given by
\begin{equation}\label{eq:JC}
J^{\mu}(q)=K^{\mu\nu}(q)A_{\nu}(q),
\end{equation}
where $q= (i\Omega_{m}, \mathbf{q})$ is a four-momentum, 
with a bosonic Matsubara frequency $i\Omega_{m}$. 
The quantity $K^{\mu\nu}$ is the EM response kernel, which is of principal
interest here. Charge conservation $(q_{\mu}J^{\mu}=0)$ 
implies that the response kernel $K^{\mu\nu}$ must satisfy
the condition $q_{\mu}K^{\mu\nu}=0$. 
The satisfaction of this condition is what we will
mean by a gauge invariant many-body theory. 

The response kernel $K^{\mu\nu}$ can be written in a general form as \cite{schrieffer}
\begin{align}\label{eq:KN}
K^{\mu\nu}(q)&=2\sum_{k}G(k_{+})\Gamma^{\mu}(k_{+},k_{-})G(k_{-})\gamma^{\nu}(k_{-},k_{+})\nonumber\\
&\quad+\frac{n}{m}\delta^{\mu\nu}(1-\delta_{0\mu}),
\end{align}
where the full and bare vertices are
$\Gamma^{\mu}(k_{+},k_{-}),$ $\gamma^{\mu}(k_{+},k_{-})$ respectively,
and $k_{\pm}\equiv k\pm q/2$ is the incoming (+) or outgoing $(-)$ momenta of a vertex.
The particle number is $n$ and $m$ denotes the fermion mass. 
The full Green's function is denoted by $G(k)$, which we define in terms of the bare 
Green's function, $G_{0}^{-1}(k)=i\omega-\xi_{\mb{k}}$, in Eq. (\ref{eq:GFE}).
Here the single particle dispersion is $\xi_{\mb{k}}=k^2/2m-\mu$,
where $\mu$ is the chemical potential.

We now introduce a framework that
encapsulates both BCS theory and
stronger correlations beyond BCS theory. 
To understand what is meant by these correlation
effects, here we consider a correlated self-energy $\Sigma_{\corr}(k)$.
In order to simultaneously describe a wide variety of theories, we define
the partially dressed Green's function
\begin{equation}\label{eq:GOA}
\left(G_{0}^{\alpha}\right)^{-1}(k)=G_{0}^{-1}(k)-\alpha\Sigma_{\corr}(k).
\end{equation}
This depends on the strong correlation contribution to
the self-energy $\Sigma_{\corr}$ for $\alpha=1$,
and does not include strong correlation effects for $\alpha=0$.
The fermionic Green's function is then
given by Dyson's equation
\begin{equation}\label{eq:GFE}
G^{-1}(k)=G_{0}^{-1}(k)-\Sigma(k),
\end{equation}
where the self-energy consists of two terms:
\begin{equation}\label{eq:SSE}
\Sigma(k)=\Sigma_{\corr}(k)-|\Delta_{\sc}|^{2}G^{\alpha}_{0}(-k),
\end{equation}
for a superconducting order parameter $\Delta_{\sc}$.
Equivalently, $\Sigma(k)=\Sigma_{\corr}(k) + \Sigma_{\sc}(k)$,
where $\Sigma_{\sc}(k) = -|\Delta_{\sc}|^{2}G^{\alpha}_{0}(-k)$ is the superconducting self-energy.

Finally, the gap equation can be written \cite{YRZ,ourreview}
as $1-g\sum_{k}G^{\alpha}_{0}(-k)G(k)=0.$ 
Multiplying both sides of this equation by $\Delta_{\sc}$, we obtain
\begin{equation}
\label{eq:GAP}
\Delta_{\sc}/g=\sum_{k}\Delta_{\sc}G^{\alpha}_{0}(-k)G(k)\equiv\sum_{k}F_{\sc}(k).\end{equation}
In this expression the anomalous Green's function $F_{\sc}(k)$ has 
dependence on $\Sigma_{\corr}(k)$ via $G_{0}^{\alpha}(k)$ and $G(k)$, 
and there is also implicit 
dependence on $\alpha$ through $G_{0}^{\alpha}(k)$.

This represents a fairly generic class of 
strongly correlated superfluid systems. When 
$\Sigma_{\corr}=0$
the system reverts to the conventional BCS theory.
Thus, the challenge is to include the correlation
effects associated with the self-energy $\Sigma_{\corr}$.
Models of this sort are associated with the work of Yang, Rice, and Zhang \cite{YRZ},
and also with the work of Refs. \cite{StrinatiPRL, StrinatiPRB, StrinatiPRB2}, 
who address BCS-BEC crossover effects via a $t$-matrix.
Also belonging to this class is an alternate $t$-matrix theory 
of BCS-BEC crossover \cite{ourreview, Hao2}, which, in
contrast to the work of Ref. \cite{StrinatiPRL}, is more
directly associated with a BCS-based ground state.

\subsection{The Ward-Takahashi identity}
In order to derive the gauge invariant EM response, 
we now apply the Ward-Takahashi identity (WTI). 
For a quantum field theory with a $U(1)$ gauge symmetry the WTI 
is an exact relation between the many-body vertex function
that appears in correlation functions and the self-energy which enters in the Green's function.
Moreover, as shown in the Supplemental Material \cite{Supplement}, 
given a full vertex that satisfies the WTI, the $f$-sum-rule is satisfied 
and thus charge is conserved. 

Given the bare Green's function $G_{0}(k)$, and the full Green's function $G(k)$,
the WTI constrains the full vertex $\Gamma^{\mu}(k_{+},k_{-})$ so that
it satisfies \cite{Ryder}
\begin{align}\label{eq:WTI}
q_{\mu}\Gamma^{\mu}(k_{+},k_{-})&=G^{-1}(k_{+})-G^{-1}(k_{-}),\nonumber\\
&=q_{\mu}\gamma^{\mu}(k_{+},k_{-})+\Sigma(k_{-})-\Sigma(k_{+}).
\end{align}
The bare WTI, $q_{\mu}\gamma^{\mu}(k_{+},k_{-})=G_{0}^{-1}(k_{+})-G_{0}^{-1}(k_{-})$, 
is satisfied for a bare vertex $\gamma^{\mu}(k_{+},k_{-})=(1,\mb{k}/m)$.
Therefore, given a self-energy $\Sigma(k)$, the above equation
provides a constraint which can be used to determine the full vertex. 

The WTI is equivalent to self-consistent perturbation theory,
and allows one to compute the exact
$n$-loop full vertex, given any $n$-loop self-energy.
If the self-energy depends on the full Green's function,
then applying the WTI leads to an integral equation for the full vertex of the Bethe-Salpeter form \cite{Nozieres}.
However, if the self-energy depends on only a finite number
of bare or partially dressed Green's functions, then this integral equation terminates,
and the full vertex can be obtained exactly. This is the
situation with regard to the strong correlation approaches we consider in this paper.

We now turn to the superconducting case.
For a superconductor, where gauge invariance is ``spontaneously broken",
the presence of a condensate
below the transition temperature
leads to a more complicated formulation of the WTI.
Imposing gauge invariance in the presence of a condensate requires low energy excitations
known as collective modes.
The explicit form of the collective modes, however, must be derived from the
gap equation \cite{Hao2}.

The Ward-Takahashi identity is equivalent to requiring
that the full vertex be obtained by performing 
all possible vertex insertions into the self-energy \cite{Nambu}.
Below the transition temperature, however, 
we must account for the effect of an external (non-dynamical) vector potential $A_{\mu}$
on the self-consistency condition (Eq. (\ref{eq:GAP})). 
This necessitates the introduction of collective mode vertices $\Pi^{\mu}(q)$, $\conj{\Pi}^{\mu}(q)$ in the full vertex,
which are inserted into every location of the condensate terms $\Delta_{\sc}$, $\Delta_{\sc}^{*}$, respectively.
In the next section we discuss these collective mode vertices in greater detail.
As shown in the Supplemental Material \cite{Supplement}, 
performing all vertex insertions into the self-energy of Eq. (\ref{eq:SSE}), 
and using Eq. (\ref{eq:WTI}), then gives the full vertex: 
\begin{align}\label{eq:FV1}
\Gamma^{\mu}&(k_{+},k_{-})=\gamma^{\mu}(k_{+},k_{-})+\Lambda^{\mu}(k_{+},k_{-})\nonumber\\
&-\Delta_{\sc}^{*}\Pi^{\mu}(q)G^{\alpha}_{0}(-k_{-})
-\Delta_{\sc}\conj{\Pi}^{\mu}(q)G^{\alpha}_{0}(-k_{+})\nonumber\\
&-|\Delta_{\sc}|^2G^{\alpha}_{0}(-k_{-})G^{\alpha}_{0}(-k_{+})\times\nonumber\\
&\quad[\gamma^{\mu}(-k_{-},-k_{+})+\alpha\Lambda^{\mu}(-k_{-},-k_{+})].
\end{align}
Here we have introduced the vertex correction $\Lambda^{\mu}(k_{+},k_{-})$, 
which relates to the correlated self-energy contribution and satisfies
$q_{\mu}\Lambda^{\mu}(k_{+},k_{-})=\Sigma_{\corr}(k_{-})-\Sigma_{\corr}(k_{+})$. 
The collective mode vertices in this expression are (as yet) 
unknowns which satisfy $q_{\mu}\Pi^{\mu}(q)=2\Delta_{\sc}$,
$q_{\mu}\bar{\Pi}(q)=-2\Delta^{*}_{\sc}$.
However, by ensuring that these collective mode vertices are consistent with the gap equation, 
a unique expression for them can be obtained \cite{Hao2}. This will be outlined in the next section. 
Using these relations, along with the bare WTI, 
one can check explicitly that this full vertex satisfies the full WTI in Eq. (\ref{eq:WTI}).

By way of comparison, we note that the full vertex in Eq. (\ref{eq:FV1})
is analogous to the BCS full vertex, but with the
mapping $\gamma^{\mu}\rightarrow\gamma^{\mu}+\alpha\Lambda^{\mu},G_{0}\rightarrow G_{0}^{\alpha}$.
The many-body effect of the correlation
term $\Sigma_{\corr}$ (in the partially dressed Green function $G_{0}^{\alpha}$) is
therefore to modify both the bare vertex and the single particle Green's function
appearing in the superconducting part of the full vertex.
The expression in Eq. (\ref{eq:FV1}) is completely general, given a self-energy of the form in Eq. (\ref{eq:SSE}).

Note that the full vertex of interest corresponds only to the ``particle'' Green's function $G(k)$;
that is, it is not the vertex in Nambu representation, which also needs vertex corrections from the
charge conjugated ``hole'' Green's function $-G^{*}(k)$.
The present formalism thus allows one to compute gauge invariant quantities without working in Nambu space.
For some cases this technique can be expressed using Nambu notation. 
However, not all strongly correlated
theories are compatible with Nambu notation. 
In what follows we will illustrate how to compute the full vertex, and corresponding response kernel,
for some examples of strongly correlated superfluids.

Two important limiting cases of the full vertex in Eq. (\ref{eq:FV1}) 
can be checked against known results.
When $\Sigma_{\corr}=0$, then $\Lambda^{\mu}=0$, and 
the full vertex reduces to the known strict BCS case \cite{Hao1}.
If we set $\Delta_{\sc}=0$, then the full vertex also reduces to the 
known full vertex in the exotic normal state \cite{Rufus1, Peter1}. 

\subsection{Collective mode vertices}\label{sec:CM}
The challenge in studying strongly correlated superfluids, at all temperatures, is to 
treat the collective modes in a manner compatible with gauge invariance. 
In this section we implement a powerful method 
of obtaining the expressions for the collective mode vertices $\Pi^{\mu}(q),$ $\bar{\Pi}^{\mu}(q)$. 
Gauge invariance alone requires that
$q_{\mu}\Pi^{\mu}(q)=2\Delta_{\sc},$ $q_{\mu}\conj{\Pi}^{\mu}(q)=-2\Delta_{\sc}^{*}.$ 
The gap equation imposes a self-consistency condition on both vertices
which we will use in order to determine the explicit form of these vertices.
This gap equation is written in Eq. (\ref{eq:GAP})
and in what follows we also consider the conjugate gap equation.

\begin{figure}[h]
\centering\includegraphics[width=4.5cm,height=3.5cm,clip]{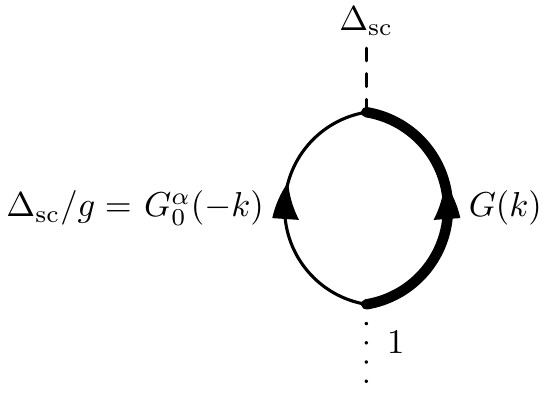}
\caption{Feynman diagram for the gap equation $\Delta_{\sc}/g=\Delta_{\sc}\sum_{k}G_{0}^{\alpha}(k)G(k)$.}
\label{fig:Gap}
\end{figure}

\begin{figure}[ht]
\centering\includegraphics[width=8.5cm,height=6cm,clip]{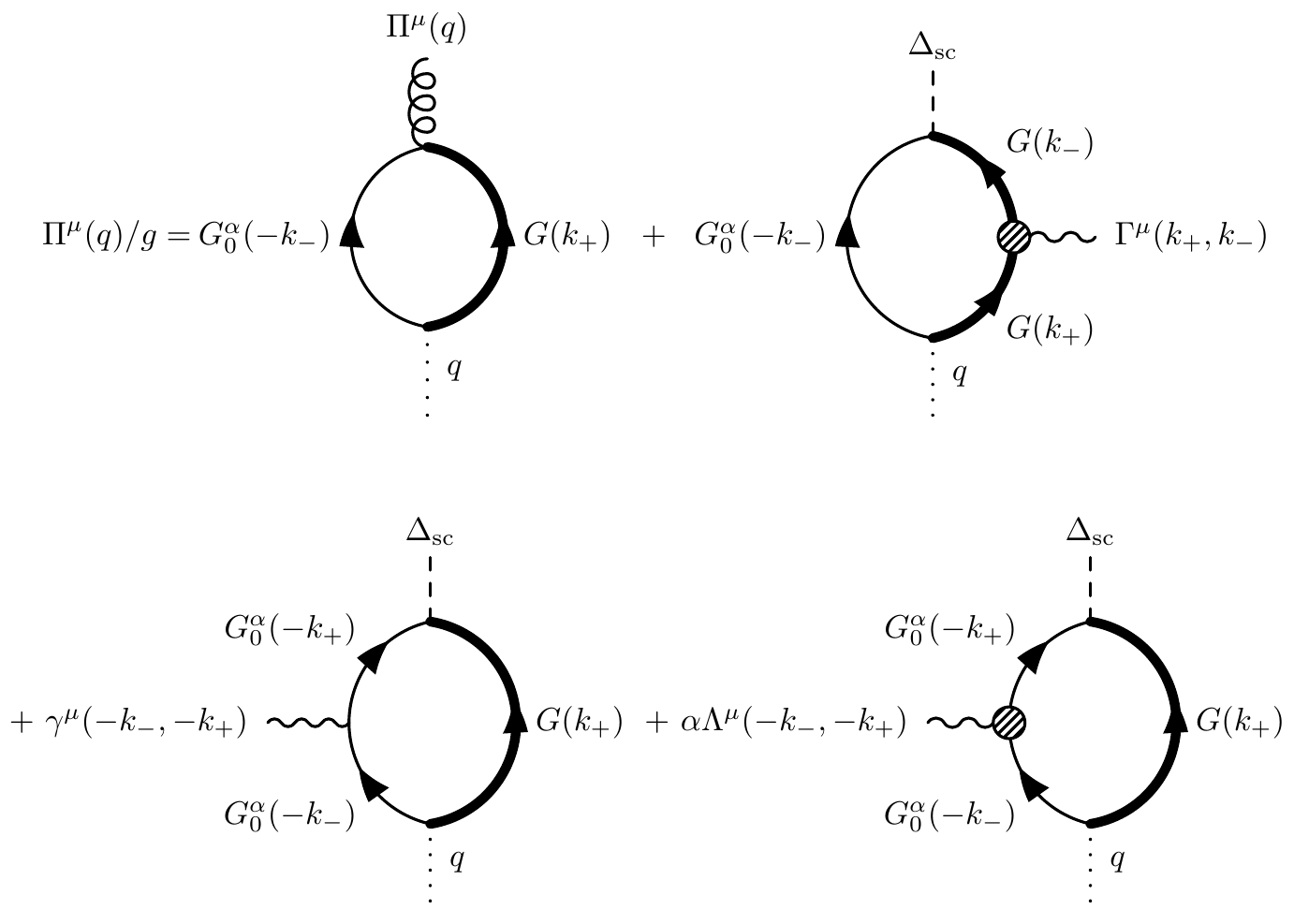}
\caption{Self consistent equation for the collective modes after performing all possible vertex insertions into the gap equation.}
\label{fig:Cmodes}
\end{figure}

In Fig. (\ref{fig:Gap}) the gap equation is expressed as a Feynman diagram. 
Diagrammatically, the collective mode vertices are obtained by performing all possible vertex insertions into the gap equation. 
In Fig. (\ref{fig:Gap}) there are three possible vertex insertions:
(1) at the $\Delta_{\sc}$ location one can insert $\Pi^{\mu}(q)$, 
(2) at the full Green function $G(k)$ location one can insert the full vertex $\Gamma^{\mu}(k_{+},k_{-})$,
(3) at the partially dressed Green function $G_{0}^{\alpha}(-k)$ location one can insert
the partially dressed vertex $\gamma^\mu(-k_{-},-k_{+})+\alpha\Lambda^{\mu}(-k_{-},-k_{+})$.
After performing these vertex insertions we obtain the equation in Fig. (\ref{fig:Cmodes}) expressed in terms of Feynman diagrams.

Mathematically, Fig. (\ref{fig:Cmodes}) implies that the collective mode vertices must satisfy the following equation
\begin{align}\label{eq:CME}
\Pi^{\mu}(q)/g&=\Pi^{\mu}(q)\sum_{k}G_{0}^{\alpha}(-k_{-})G(k_{+})\nonumber\\
& +\Delta_{\sc}\sum_{k}G_{0}^{\alpha}(-k_{-})G(k_{+})\Gamma^{\mu}(k_{+},k_{-})G(k_{-})\nonumber\\
& +\Delta_{\sc}\sum_{k}\biggl(G_{0}^{\alpha}(-k_{-})G_{0}^{\alpha}(-k_{+})G(k_{+})\nonumber\\
&\quad\quad\quad\times\left[\gamma^{\mu}(k_{+},k_{-})+\Lambda^{\mu}(k_{+},k_{-})\right]\biggr).
\end{align}
Notice that the full vertex $\Gamma^{\mu}(k_{+},k_{-})$ appears in this expression. 
The full vertex was already determined in Eq. (\ref{eq:FV1}) using the Ward-Takahashi identity. 
Therefore if we insert the expression for the full vertex, which contains the collective mode vertices, 
into Eq. (\ref{eq:CME}) (and its conjugate), then Eq. (\ref{eq:CME}) (and its conjugate)
becomes a self-consistent set of equations for the collective mode vertices $\Pi^{\mu}$ and $\bar{\Pi}^{\mu}$.
The solution to this self-consistent set of linear equations will uniquely determine the collective mode vertices.

Inserting the full vertex into Eq. (\ref{eq:CME}), and doing the same analysis for the conjugate gap equation,
then gives the following two self-consistent equations for the collective mode vertices
\begin{align}\label{eq:LV1}
\Pi^{\mu}(q)/g&=\Pi^{\mu}(q)\sum_{k}G(k_{+})G^{\alpha}_{0}(-k_{-})\left[1-\Delta_{\sc}^{*}F_{\sc}(k_{-})\right]\nonumber\\
&-\conj{\Pi}^{\mu}(q)\sum_{k}F_{\sc}(k_{+})F_{\sc}(k_{-})\nonumber\\
&+\sum_{k}\left[\gamma^{\mu}(k_{+},k_{-})+\Lambda^{\mu}(k_{+},k_{-})\right]G(k_{+})F_{\sc}(k_{-})\nonumber\\
&+\sum_{k}\biggl(\left[\gamma^{\mu}(-k_{-},-k_{+})
+\alpha\Lambda^{\mu}(-k_{-},-k_{+})\right]\nonumber\\
&\quad\times F_{\sc}(k_{+})G^{\alpha}_{0}(-k_{-})\left[1-\Delta^{*}_{\sc}F_{\sc}(k_{-})\right]\biggr),\\
\conj{\Pi}^{\mu}(q)/g&=\conj{\Pi}^{\mu}(q)\sum_{k}G(k_{-})G^{\alpha}_{0}(-k_{+})\left[1-\Delta_{\sc}F^{*}_{\sc}(k_{+})\right]\nonumber\\
&-\Pi^{\mu}(q)\sum_{k}F^{*}_{\sc}(k_{+})F^{*}_{\sc}(k_{-})\nonumber\\
&+\sum_{k}\left[\gamma^{\mu}(k_{+},k_{-})+\Lambda^{\mu}(k_{+},k_{-})\right]F^{*}_{\sc}(k_{+})G(k_{-})\nonumber\\
&+\sum_{k}\biggl(\left[\gamma^{\mu}(-k_{-},-k_{+})
+\alpha\Lambda^{\mu}(-k_{-},-k_{+})\right]\nonumber\\
&\quad\times G^{\alpha}_{0}(-k_{+})F^{*}_{\sc}(k_{-})\left[1-\Delta_{\sc}F^{*}_{\sc}(k_{+})\right]\biggr).
\label{eq:LV2}\end{align}

This is conveniently expressed as a matrix equation if we define the two-point correlation functions
\begin{align}
Q_{+-}(q)&=1/g-\sum_{k}G(k_{+})G^{\alpha}_{0}(-k_{-})\left[1-\Delta_{\sc}^{*}F_{\sc}(k_{-})\right],\nonumber\\
Q_{++}(q)&=\sum_{k}F_{\sc}(k_{+})F_{\sc}(k_{-}),\nonumber\\
P^{\mu}_{+}(q)&=\sum_{k}\left[\gamma^{\mu}(k_{+},k_{-})+\Lambda^{\mu}(k_{+},k_{-})\right]G(k_{+})F_{\sc}(k_{-})\nonumber\\
&+\sum_{k}\left[\gamma^{\mu}(-k_{-},-k_{+})+\alpha\Lambda^{\mu}(-k_{-},-k_{+})\right]\times\nonumber\\
&\quad F_{\sc}(k_{+})G^{\alpha}_{0}(-k_{-})\left[1-\Delta^{*}_{\sc}F_{\sc}(k_{-})\right],
\end{align}
and $Q_{-+}(q)=Q_{+-}^{*}(q)$, $Q_{--}(q)=Q_{++}^{*}(q)$, $\Delta_{\sc}^{*}P^{\mu}_{+}(q)=\Delta_{\sc}P^{\mu}_{-}(-q)$.
To connect to the literature, we define an alternative set of of two-point correlation functions $Q_{ab}$ and $Q^{a\mu}$, where $a,b=1,2$ through, 
$Q_{11}=Q_{+-}+Q_{-+}+Q_{++}+Q_{--},$ $Q_{22}=Q_{+-}+Q_{-+}-Q_{++}-Q_{--},$ $Q_{12}=i(Q_{+-}-Q_{-+}+Q_{--}-Q_{++}),$
$Q_{21}=-i(Q_{+-}-Q_{-+}+Q_{++}-Q_{--})$,
and $Q^{1\mu}=-\left(P^{\mu}_{+}+P^{\mu}_{-}\right)$, $Q^{2\mu}=-i(P^{\mu}_{-}-P^{\mu}_{+})$.
Similarly, we define the collective mode vertices $\Pi^{\mu}_{1,2}(q)$ through
$\Pi^{\mu}(q)=\Pi^{\mu}_{1}(q)+i\Pi^{\mu}_{2}(q)$, $\bar{\Pi}^{\mu}(q)=\Pi^{\mu}_{1}(q)-i\Pi^{\mu}_{2}(q)$.
This amounts to a change of basis from a complex to a real and imaginary parameterization.
From Eq. (\ref{eq:LV1}) and Eq. (\ref{eq:LV2}), these vertices satisfy the relation
\begin{equation}\label{eq:CM}
\left(\begin{array}{c}\Pi^{\mu}_{1}\\ \Pi^{\mu}_{2}\end{array}\right)=
-\left(\begin{array}{cc}Q_{11}&Q_{12}\\Q_{21}&Q_{22}\end{array}\right)^{-1}\left(\begin{array}{c}Q^{1\mu}\\Q^{2\mu}\end{array}\right).
\end{equation}
The form of these collective mode vertices is structurally similar to the BCS case \cite{Hao1, Hao2}, 
and in the strict BCS limit they agree with the literature \cite{Hao1}.
The matrix $Q_{ab}$ can be interpreted as a propagator for bosonic degrees of freedom. 
However, the explicit response functions entering on the right hand side of Eq. (\ref{eq:CM}) are
modified due to the presence of the self-energy $\Sigma_{\corr}$. 

In the Supplemental Material \cite{Supplement} we
verify that the collective mode vertices $\Pi^{\mu}(q)$ and $\bar{\Pi}^{\mu}(q)$ satisfy the gauge invariant conditions 
$q_{\mu}\Pi^{\mu}(q)=2\Delta_{\sc}$, $q_{\mu}\conj{\Pi}^{\mu}(q)=-2\Delta_{\sc}^{*}$, which was assumed in their definitions.

\subsection{Vertex correction \texorpdfstring{$\Lambda^{\mu}$}{L}}
We can now summarize the central results of this paper, and repeat key equations. 
The full electromagnetic response kernel can generically be written as
\begin{align}
K^{\mu\nu}(q)&=2\sum_{k}G(k_{+})\Gamma^{\mu}(k_{+},k_{-})G(k_{-})\gamma^{\nu}(k_{-},k_{+})\nonumber\\
&\quad+\frac{n}{m}\delta^{\mu\nu}(1-\delta_{0\mu}),\nonumber\tag{\ref{eq:KN}}
\end{align}
where the full vertex
\begin{align}
\Gamma^{\mu}&(k_{+},k_{-})=\gamma^{\mu}(k_{+},k_{-})+\Lambda^{\mu}(k_{+},k_{-})\nonumber\\
&-\Delta_{\sc}^{*}\Pi^{\mu}(q)G^{\alpha}_{0}(-k_{-})
-\Delta_{\sc}\conj{\Pi}^{\mu}(q)G^{\alpha}_{0}(-k_{+})\nonumber\\
&-|\Delta_{\sc}|^2G^{\alpha}_{0}(-k_{-})G^{\alpha}_{0}(-k_{+})\times\nonumber\\
&\quad[\gamma^{\mu}(-k_{-},-k_{+})+\alpha\Lambda^{\mu}(-k_{-},-k_{+})],\nonumber\tag{\ref{eq:FV1}}
\end{align}
contains contributions due to both the collective mode vertices $\Pi^{\mu}$ and $\bar{\Pi}^{\mu}$ 
(computed in Eq. (\ref{eq:CM})) and the 
vertex contribution $\Lambda^{\mu}$ arising from the self-energy $\Sigma_{\corr}$.

The techniques described above are sufficient to calculate a gauge invariant response
function for a large class of theories. 
All that is required to derive the full gauge invariant electromagnetic
response is to arrive at a form of $\Lambda^{\mu}$. 
This vertex depends on the details of the correlation self-energy $\Sigma_{\corr}$, so 
we must consider it on a case by case basis. We now consider three relevant examples from the literature.

\subsubsection{Pairing pseudogap}
The first type of strong correlations we study is that proposed in Ref. \cite{Norman}
at a phenomenological level and in Ref. \cite{ourreview} from a more microscopic perspective.
In Ref. \cite{Kosztin} an early attempt to address how collective modes are affected
by these pseudogap effects was performed.
This model is based on a BCS like self-energy but with a normal state
gap $\Delta_{\pg}$. For this model, which we call the ``pairing pseudogap approximation",
$\alpha=0$ in Eq. (\ref{eq:GOA}), and the correlated self-energy in Eq. (\ref{eq:SSE})
is given by 
\begin{equation}
\Sigma_{\corr}(k)=-\Delta_{\pg}^2G_{0}(-k).
\end{equation}
The pairing gap $\Delta_{\pg}$ is non-zero in the range of temperatures $T^{*}>T_{c}>0$,
where $T^{*}$ is the mean-field transition temperature ($\Delta_{\pg}(T^{*}=0$)). 
At a more microscopic level \cite{ourreview}
$\Delta_{\pg}$ is to be associated with non-condensed (finite momentum)
pairs and is distinct from the superconducting order parameter $\Delta_{\sc}$ which
corresponds to a condensate of pairs at zero net momentum.

Unlike the order parameter $\Delta_{\sc}$,
the gap $\Delta_{\pg}$ does not fluctuate in the presence of $A_\mu$.
Nevertheless, its inclusion in the self-energy will lead to a vertex correction.
Using this form of $\Sigma_{\corr}(k)$, along with the definition 
$q_{\mu}\Lambda^{\mu}(k_{+},k_{-})=\Sigma_{\corr}(k_{-})-\Sigma_{\corr}(k_{+})$, 
we obtain
\begin{equation}
\Lambda^{\mu}(k_{+},k_{-})=\Delta_{\pg}^2G_{0}(-k_{-})\gamma^{\mu}(-k_{-},-k_{+})G_{0}(-k_{+}).\label{eq:Lmu-pg}
\end{equation}
Inserting this expression into Eq. (\ref{eq:FV1}), along with $\alpha=0$, 
then gives the full superconducting vertex in the
pseudogap approximation. 

Note that the pseudogap self-energy is an approximation of a theory with 
$\alpha=0$ and $\Sigma_{\corr}(k)=\sum_{l}t_{\pg}(l)G_{0}(l-k)$, 
where $t_{\pg}(l)$ is a $t$-matrix. This theory was considered in Ref.~\cite{Hao2}, and the vertex $\Lambda^{\mu}$ was 
calculated exactly.
The exact $t_{\pg}$ depends on the full Green's function, so the exact $\Lambda^{\mu}$ 
will itself depend on the full vertex $\Gamma^{\mu}$, 
and thus a self-consistent integral equation will arise for $\Gamma^{\mu}$. 
In the pairing pseudogap approximation, $\Delta_{\pg}$ is constructed such that it contains
no external momentum. Thus no vertex insertions into the gap are possible in $\Lambda^{\mu}$,
resulting in the above condition that $\Delta_{\pg}$ does not fluctuate with $A_\mu$.

\subsubsection{YRZ model}
As a second model we consider a phenomenological self-energy
developed for the high-$T_{c}$ superconductors and
associated with Yang, Rice, and Zhang \cite{YRZ}. 
This is known as the YRZ model.
For the YRZ model, in Eq. (\ref{eq:GOA}) and Eq. (\ref{eq:SSE}) one sets  $\alpha=1$ and 
\begin{equation}
\Sigma_{\corr}(k)=-\Delta_{\pg}^2G_{0}(-k).
\end{equation}
Since $\Sigma_{\corr}(k)$ is the same as in the pairing pseudogap approximation,
in the YRZ model we also obtain 
\begin{equation}
\Lambda^{\mu}(k_{+},k_{-})=\Delta_{\pg}^2G_{0}(-k_{-})\gamma^{\mu}(-k_{-},-k_{+})G_{0}(-k_{+}).
\end{equation}
Inserting this vertex correction into Eq. (\ref{eq:FV1}), along with $\alpha=1$,
then gives the full superconducting vertex in the YRZ model.
In the normal state, this full vertex, along with the response kernel in Eq. (\ref{eq:KN}),
is in agreement with the results obtained in Ref. \cite{Peter1}. 
Here we have extended this work to the superconducting case.

\subsubsection{Particle-only \texorpdfstring{$t$}{t}-matrix}
A third and final model was introduced by Strinati and co-workers using
a generalized $t$-matrix
\cite{StrinatiPRL, StrinatiPRB, StrinatiPRB2}.
In this model the self-energy
is obtained from Eq. (\ref{eq:GOA}) and Eq. (\ref{eq:SSE})
by setting $\alpha=1$ and 
\begin{equation}\label{eq:TMSE}
\Sigma_{\corr}(k)=\sum_{l}t(l)G_{\BCS}(l-k).
\end{equation}
Here $G_{\BCS}$ is the full Green's function as would be defined in a pure BCS theory;
$t(l)$ is a $t$-matrix, the details of which are presented in the Supplemental Material \cite{Supplement}.
In Ref. \cite{StrinatiPRB2}, the authors propose ``good candidates" for
the response function Feynman diagrams.
Here we emphasize that the WTI provides a direct procedure to determine 
not just good candidates but the exact full vertex,
given in Eq. (\ref{eq:FV1}), which is manifestly gauge invariant.
The challenge here is in determining the exact vertex correction $\Lambda^{\mu}(k_{+},k_{-})$.
This is more complicated than in the previous two cases.
Nevertheless, following the procedure outlined above,
the vertex correction due to this self-energy can be obtained
by performing all possible vertex insertions into all internal lines.
That is, by inserting all possible vertices into both the Green's function and into the $t$-matrix.
In the Supplemental Material \cite{Supplement} we explicitly derive the vertex correction
$\Lambda^{\mu}$ for the self-energy appearing in Eq. (\ref{eq:TMSE}).
We should note that the authors of this body of work do not presume a
self-consistent gap equation, such as that appearing in Eq. (\ref{eq:GAP}),
and such as we have assumed in arriving at Eq. (\ref{eq:CM}).
Rather, they fix the order parameter to be the same as in BCS theory. 

In summary, this section has shown how to derive a gauge invariant full vertex
for a generic self-energy of the form in Eq. (\ref{eq:SSE}).
Using the Ward-Takahashi identity there is an exact procedure to determine the full vertex.
Moreover there is an analogous procedure to determine the collective modes
and thus maintain gauge invariance.
The resulting Feynman diagrams, which are shown in the 
Supplemental Material \cite{Supplement}, are then completely determined.

\section{Alternative scheme to Ward-Takahashi: Path Integral}
\subsection{Gauge invariant electrodynamics}\label{sec:PathIntegral}
A large class of theories in the literature derive the gauge invariant
electromagnetic response using a path integral approach \cite{Altland, Kitagawa, Yakovenko}. 
We now connect, when possible, the above results using the Ward-Takahashi identity to
the EM response as calculated in the path integral literature.
Here we will include both amplitude and phase fluctuations of the order parameter~\cite{Taylor, Randeria1}.
This is in contrast to previous studies~\cite{Altland, Kitagawa, Yakovenko} which incorporate only phase fluctuations.
We introduce these amplitude fluctuations in large part in order to address the compressibility sum-rule. 

The inverse Nambu Green's function is 
$\mathcal{G}^{-1}=\mathcal{G}^{-1}_{0}-\Sigma$, where
$\mathcal{G}_{0}^{-1}=i\omega-\xi_{\mb{k}}\tau_{3}$ and
the self-energy is $\Sigma=-\Delta(x)\tau_{+}-\Delta^{*}(x)\tau_{-}$.
The Nambu Pauli matrices are $\tau_{1,2,3}$, which define the
raising and lowering operators $\tau_\pm=\frac{1}{2}\left(\tau_1 \pm i \tau_2\right)$.
We begin with the action functional in terms of the Hubbard-Stratonovich field $\Delta$ \cite{Taylor}:
\begin{equation}
S[\Delta^{*},\Delta, A^{\mu}]=-\mathrm{Tr\ ln}\left[-\mathcal{G}^{-1}\right]+\int dx\ \frac{|\Delta(x)|^2}{g},
\end{equation}
and following convention, the trace $\textrm{Tr}$ represents a trace over both Nambu and position indices.
We now follow the literature and perform the saddle point expansion. To lowest order the effective action is
$S_{\eff}[\Delta^{*},\Delta,A^{\mu}]=S_{\mf}[\Delta_{\mf}^{*},\Delta_{\mf}]$,
where the mean-field (mf) action is
\begin{equation}\label{eq:MFA}
S_{\mf}[\Delta_{\mf}^{*},\Delta_{\mf}]=-\Tr\ \mathrm{ln}\left[-\mathcal{G}_{\mf}^{-1}\right]+\int dx\ \frac{|\Delta_{\mf}|^2}{g},
\end{equation}
and the inverse mean-field Nambu Green's function is
$\mathcal{G}^{-1}_{\mf}=\mathcal{G}^{-1}_{0}-\Sigma[\Delta(x)\rightarrow\Delta_{\mf}]$.
The BCS gap equation then follows upon setting $\delta S_{\mf}[\Delta_{\mf}^{*},\Delta_{\mf}]/\delta\Delta_{\mf}^{*}=0$.
It is straightforward to see that the resulting response kernel is not gauge invariant.

We now calculate the gauge invariant EM response kernel $K^{\mu\nu}$.
In order to implement gauge invariance, the conventional literature
introduces fluctuations $\eta(x)$ about the mean-field 
value of the order parameter $\Delta_{\mf}$, expressing 
$\Delta(x)=\Delta_{\mf}+\eta(x)$. 
(In Sec. (\ref{sec:Ward}), $\Delta_{\sc}\equiv\Delta_{\mf}$ for strict BCS theory.) 
Expanding the action functional to second order in $\eta(x)$ gives
$S[\Delta^{*},\Delta,A^{\mu}] \approx S_{\mf}[\Delta_{\mf}^{*},\Delta_{\mf}]+S^{(2)}[\eta^{*},\eta,A^{\mu}]$.
To calculate $S^{(2)}[\eta^{*},\eta,A^{\mu}]$, 
we first consider fluctuations of the Green's function about the mean-field solution:
\begin{equation}
\mathcal{G}^{-1}-\mathcal{G}^{-1}_{\mf}=-\delta\Gamma-\Sigma_{\eta},
\end{equation}
where $\delta\Gamma=\Gamma_{1}+\Gamma_{2}$, with $\Gamma_{1}=\gamma_{\mu}A^{\mu}$, $\Gamma_{2}=(\mb{A}^2/2m)\tau_{3}$, is a vector potential fluctuation 
and $\Sigma_{\eta}=\Sigma[\Delta(x)\rightarrow\eta(x)]$ is a gap fluctuation.
Expanding to second order in $\eta$ and $A_{\mu}$, the second order action functional is
\begin{align}
&S^{(2)}[\eta^{*},\eta,A^{\mu}]\nonumber\\
&=\frac{1}{2}\sum_{q}\left[ A_{\mu}(q)K^{\mu\nu}_{0,\mf}(q)A_{\nu}(-q)+\eta_a(q) Q^{ab}_{\mf}(q) \eta_b(-q)\right] \nonumber\\
&+\frac{1}{2}\sum_{q}\left[A_{\mu}(q) Q^{\mu b}_{\mf}(q) \eta_b(-q) + \eta_a(q) Q^{a\nu}_{\mf}(q)A_{\nu}(-q)\right]\nonumber.
\end{align}
In this expression we write $\eta(x)=\eta_{1}(x)-i\eta_{2}(x)$ with $\eta_{1}(x),\eta_{2}(x)\in\mathbb{R}$.
This decomposes the fluctuations into their (Cartesian) real and imaginary parts, which amounts to an amplitude and phase decomposition.
Since we keep the saddle point condition at the mean-field level, an explicit amplitude and phase decomposition,
in polar coordinates, will lead to the same electromagnetic response. 
(If one uses a different saddle point condition, not relevant to this work,
then issues associated with the use of either a Cartesian or polar decomposition may arise \cite{Randeria1}.) 
Even within this framework, we shall point out an inconsistency within the conventional path integral formalism
in failing to satisfy the compressibility sum-rule.

To complete the calculation, we transform to momentum space, $k=(i\omega_{n},\mb{k})$ and $q=(i\Omega_{m},\mb{q})$, where 
$i\omega_{n}$ $(i\Omega_{m})$ is a fermionic (bosonic) Matsubara frequency. If we denote
the trace over Nambu indices by $\textrm{tr}$, then the ``bubble" response kernel is
$K^{\mu\nu}_{0,\mf}(q)=\tr\sum_{k}\mathcal{G}_{\mf}(k_{+})\gamma^{\mu}(k_{+},k_{-})\mathcal{G}_{\mf}(k_{-})\gamma^{\nu}(k_{-},k_{+})
+\tfrac{n}{m}\delta^{\mu\nu}(1-\delta_{\mu0})$
and the two-point response function 
$Q^{ab}_{\mf}(q)=\tfrac{2}{g}\delta_{ab}+\tr\sum_{k}\mathcal{G}_{\mf}(k_{+})
\tau_{a}\mathcal{G}_{\mf}(k_{-})\tau_{b}$ can be viewed as a bosonic propagator. 
We also have $Q^{\mu a}_{\mf}(q)=-\tr\sum_{k}\mathcal{G}_{\mf}(k_{+})\gamma^{\mu}(k_{+},k_{-})\mathcal{G}_{\mf}(k_{-})\tau_{a}$, and
$Q^{b\nu}_{\mf}(q)$ has $(\mu,a)\leftrightarrow(b,\nu)$.
These mean-field response functions are equivalent
to previous results in the literature \cite{Hao1}. 
They are also equivalent to the response functions which appear in
Eq.~(\ref{eq:CM}) for a theory with only a strict BCS self-energy.

After integrating out the $\eta$ field, the beyond-mean-field effective action contribution is given by
\begin{align}\label{eq:SFL}
S_{\eff}-S_{\mf}&=\sum_{q} A_{\mu}(q)K^{\mu\nu}_{\mf}(q)A_{\nu}(-q)\nonumber\\
&\quad +\frac{1}{2}\mathrm{Tr}\ \mathrm{ln}\left[Q^{ab}_{\mf}(q)\right].
\end{align}
Thus the fluctuation action decomposes into two separate terms.
The second term in the fluctuation action provides a contribution to
thermodynamics arising from Gaussian fluctuations. 
This form of the Gaussian fluctuation part of the action
is equivalent to the standard results in the literature \cite{Randeria1}.
The first term is the gauge invariant EM response kernel,
with both amplitude and 
phase fluctuations of the order parameter included,
defined by $K^{\mu\nu}_{\mf}(q)=K^{\mu\nu}_{0,\mf}(q)-
\sum_{a,b}Q^{\mu a}_{\mf}(q)\left[Q^{ab}_{\mf}(q)\right]^{-1}Q^{b\nu}_{\mf}(-q)$. 
If we expand the response kernel appearing in Eq. (\ref{eq:SFL}), 
then we obtain \cite{Hao1, Kulik}:
\begin{widetext}
\begin{equation}\label{eq:KRS}
K^{\mu\nu}_{\mf}=K^{\mu\nu}_{0,\mf}-\frac{Q_{11} Q^{\mu 2}_{\mf} Q^{2\nu}_{\mf}+Q_{22}Q^{\mu 1}_{\mf}Q^{1\nu}_{\mf}
-Q_{12}Q^{\mu 1}_{\mf}Q^{2\nu}_{\mf}-Q_{21}Q^{\mu 2}_{\mf}Q^{1\nu}_{\mf}}{Q_{11}Q_{22}-Q_{12}Q_{21}}.
\end{equation}
\end{widetext}
In Ref. \cite{Hao1} it is proved that 
the response kernel in Eq. (\ref{eq:KRS}) is both gauge invariant
$q_{\mu}K^{\mu\nu}_{\mf}(q)=0$, and charge conserving $K^{\mu\nu}_{\mf}(q)q_{\nu}=0$.
References \cite{Hao1,Kulik} used a matrix linear response formalism known as
``consistent fluctuation of the order parameter". Our derivation, however, 
is based on the path integral. 

\subsection{Inconsistency with the compressibility sum-rule}
We now turn to the implications of the
two contributions to the action in 
Eq. (\ref{eq:SFL}).
Here we focus on the compressibility sum-rule, 
which provides an important consistency check on the path integral approach.
The explicit form of the compressibility sum-rule is \cite{Pines}:
\begin{equation}\label{eq:CSR}
\mathrm{lim}_{\mb{q}\rightarrow0}\left[K^{00}(\omega=0,\mb{q})\right]=-\frac{\p n}{\p\mu}.
\end{equation}
This sum-rule
shows how to associate the electromagnetic contributions
to the action with their counterpart
contributions to the thermodynamic response.

The compressibility, $\kappa=n^{-2}(\p n/\p\mu)$,
is then related to the density response via Eq. (\ref{eq:CSR}).
Here the real frequency $\omega$ is the analytic continuation of the Matsubara frequency $i\Omega$,
defined by $i\Omega=\omega+i\gamma$ with $\gamma \rightarrow 0$.
The relationship in Eq. (\ref{eq:CSR}) is particularly useful in
characterizing various orders of approximation within the path integral scheme.
This is because at the heart of the path integral is
a close connection between electrodynamics and thermodynamics.
With the inclusion of amplitude fluctuations, which are essential for this sum-rule,
we can now test the compressibility sum-rule within the
standard path integral formalism in the literature.

Note that, this sum-rule depends on the number equation.
Consistency would seem to require that we include Gaussian
fluctuations $n_{\fl}=-\beta^{-1} \p S_{\fl}[\Delta_{\mf}^{*},\Delta_{\mf}]/\p\mu$
to the number equation coming from the second line in Eq.~(\ref{eq:SFL}).
This is, in fact, incorrect and points to an underlying inconsistency.
Instead, we will show the proper calculation level for thermodynamics is
that of pure mean-field, giving a mean-field particle number
\begin{equation}
n_{\mf}=-\frac{1}{\beta} \frac{\p S_{\mf}[\Delta_{\mf}^{*},\Delta_{\mf}]}{\p\mu}=2\sum_{k}G(k).
\end{equation}

Taking the derivative of the mean-field
number equation with respect to $\mu$ gives
\begin{equation}\label{eq:dndu0}
\frac{\p n_{\mf}}{\p\mu}=-2\sum_{k}\left[G^{2}(k)-F^2(k)+2G(k)F(k)\frac{\p\Delta_{\mf}}{\p\mu}\right],
\end{equation}
where we henceforth take $\Delta_{\mf}=\Delta_{\mf}^{*}$ for convenience. 
Here we define the single particle Green's function in terms of
the Nambu Green's function
by $G(k)=(\mathcal{G}_{\mf}(k))_{11}=-(\mathcal{G}_{\mf}(-k))_{22}$, and the anomalous Green's function is similarly
$F(k)=\Delta_{\mf} G(k)G_{0}(-k)=(\mathcal{G}_{\mf}(k))_{12}=(\mathcal{G}^{*}_{\mf}(k))_{21}$.
The fluctuation of the mean-field gap with respect to the chemical potential, $\p \Delta_{\mf} / \p \mu$, can
be found using the BCS gap equation 
\begin{equation}\label{eq:GPE}
\mathrm{GAP}[\Delta_{\mf}, \mu] := \frac{\Delta_{\mf}}{g} - \sum_{k}\mathrm{Tr}[\mathcal{G}(k)\tau_{-}]=0.
\end{equation}
Since $\Delta_{\mf}$ depends on $\mu$, by taking the total derivative with respect to $\mu$, 
we arrive at the condition
\begin{equation}
\frac{\p\Delta_{\mf}}{\p\mu} = - \frac{\p\mathrm{GAP}/\p\mu}{\p\mathrm{GAP}/\p\Delta_{\mf}}.
\end{equation}

To see that the compressibility sum-rule is satisfied, 
notice that $\p\mathrm{GAP}/\p\mu = 2\sum_{k}G(k)F(k)$ and
$\p\mathrm{GAP}/\p\Delta=2\sum_{k}F(k)F(k)$. 
Therefore, the last term in Eq.~(\ref{eq:dndu0}) can be expressed as 
$2\frac{\left(\p\mathrm{GAP}/\p\mu\right)^2}{\p\mathrm{GAP}/\p\Delta_{\mf}}$.
Now, in the limit that $\omega=0, \mb{q}\rightarrow0$, 
the following identifications can be made:
$Q^{10}_{\mf}=2\p\mathrm{GAP}/\p\mu$, 
and $Q^{11}_{\mf}=2\p\mathrm{GAP}/\p\Delta_{\mf}$. 
By computing the summation over Matsubara frequencies, 
one also obtains $2\sum_{k}\left[G^{2}(k)-F^{2}(k)\right]=K^{00}_{0,\mf}$. 

Therefore, using Eq. (\ref{eq:KRS}), Eq. (\ref{eq:dndu0}) now becomes
\begin{equation}
-\frac{\p n_{\mf}}{\p\mu}=K^{00}_{0,\mf}-\frac{Q^{10}_{\mf}Q^{01}_{\mf}}{Q_{11}}=
K^{00}(0,\mb{q}\rightarrow0).
\end{equation}
This demonstrates the expected consistency between
$-(\p n_{\mf}/\p\mu)$ and $K^{00}(0,\mb{q}\rightarrow0)$
and proves the compressibility sum-rule at the BCS level \cite{Hao1}.

The reason for the need to include amplitude fluctuations in the density-density response
can be seen from Eq. (\ref{eq:dndu0}). This equation shows that fluctuations in the gap ($\p\Delta_{\mf}/\p\mu$) 
must be included, and therefore amplitude fluctuations in the gap are necessary in order to satisfy 
the compressibility sum-rule. If only phase fluctuations are retained, the compressibility sum-rule is violated.
For a different context where amplitude fluctuations are important see Ref. \cite{Salasnich}.

The compressibility sum-rule has only been satisfied by 
ignoring the Gaussian fluctuations in the number equation. 
Had these been included, we would obtain 
$-\p n/\p\mu=-\p n_{\mf}/\p\mu-\p n_{\fl}/\p\mu \neq K^{00}(0,\mb{q}\rightarrow0),$
which violates the compressibility sum-rule.

In summary, the path integral formalism, as currently applied in the literature,
treats electrodynamics and thermodynamics inconsistently.
In this derivation of gauge invariant electrodynamics at the BCS level, 
beyond BCS fluctuations are necessarily incorporated in thermodynamics.
However, these thermodynamic fluctuations should not 
appear in the number equation if the compressibility sum rule is to be satisfied. 
The discussion in Sec. (\ref{sec:Ward}) provides insights into the resolution to this inconsistency:
there gauge invariance is obtained by determining
the collective modes that arise due to vertex insertions into the gap equation. 
This suggests that, within the path integral formalism,
one should consider a new saddle point condition in the presence of a non-zero vector potential. 
More details on this resolution are presented elsewhere \cite{ourcompanion}.

\section{Conclusions}
The goal of this paper was to show how to arrive at a proper
gauge invariant description of the electromagnetic response
in strongly correlated fermionic superfluids.
In this paper correlation effects are represented by 
``correlated self-energy" contributions which appear in addition to the usual
BCS self-energy of the condensate.
Using the Ward-Takahashi identity,
and adopting a rather generic class of such models
(widely used for the high temperature superconductors and ultra cold gases)
we are able to give exact expressions for the electromagnetic response.
The results appear in a closed form as a consequence
of the fact that the correlation self-energy
depends on only bare or partially dressed Green's functions.
Our method, which obtains expressions for
all vertex corrections and collective modes
in a manner compatible with the $f$-sum-rule,
is an important tool for studying
strongly correlated superfluids and superconductors.

For comparison we also discuss an
alternative tool which builds on the path integral approach.
With few exceptions 
this scheme has been used to address the
BCS-level response, i.e., in the absence of stronger correlations.
In contrast to approaches which build on the Ward-Takahashi
identity, here gauge invariance 
and the $f$-sum-rule are relatively straightforward to ensure. 
What is more complicated is to arrive at consistency
with the compressibility sum-rule.
This sum-rule relates electrodynamics and thermodynamics
and provides a natural test of the path integral
scheme, since the two are simultaneously calculated.
We show that in the conventional path integral literature for
the gauge invariant electrodynamics at the BCS level, 
the compressibility sum-rule is violated.

\textit{Acknowledgements.}$-$
This work was supported by NSF-DMR-MRSEC 1420709.
We are particularly grateful to Hao Guo and Yan He
for sharing their insights and for sending us their
preprints on a related topic.

\bibliography{Review}
\clearpage


\section*{Supplementary material}

# Supplementary Material: Gauge invariant theories of linear response for strongly correlated superconductors

Rufus Boyack, Brandon M. Anderson, Chien-Te Wu, and K. Levin
*James Franck Institute, University of Chicago, Chicago, Illinois 60637, USA*

## I. OBTAINING THE FULL VERTEX USING THE WARD-TAKAHASHI IDENTITY

Here we show how to apply the Ward-Takahashi identity to obtain the gauge invariant full vertex for a given self-energy. If we define the partially dressed Green's function $G_0^\alpha(k)$ by

$$(G_0^\alpha)^{-1}(k) = G_0^{-1}(k) - \alpha \Sigma_{\text{corr}}(k), \tag{1}$$

where $\Sigma_{\text{corr}}$ is a self-energy describing strong correlations, then the class of self-energies considered in the main text are of the form

$$\Sigma(k) = \Sigma_{\text{corr}}(k) - |\Delta_{\text{sc}}|^2 G_0^\alpha(-k). \tag{2}$$

The second term in this expression represents the superconducting self-energy $\Sigma_{\text{sc}}(k) = -|\Delta_{\text{sc}}^2|G_0^\alpha(-K)$. For convenience we treat $\Delta_{\text{sc}}$ and $\Delta_{\text{sc}}^*$ as independent degrees of freedom. This will be important in the next section, but for now it is not essential. Writing the self-energy in this form shows that the second term is a BCS like self-energy, but with the bare Green's function $G_0$ replaced by the partially dressed Green's function $G_0^\alpha$. Strict BCS theory is obtained by setting $\Sigma_{\text{corr}} = 0$. The three models that we will consider are the pairing pseudogap approximation [1, 2], the Yang, Rice, and Zhang (YRZ) model [3], and the $t$-matrix model of Ref. [4]. For the pairing pseudogap approximation, $\alpha = 0, \Sigma_{\text{corr}}(k) = -\Delta_{\text{pg}}^2 G_0(-k)$, for the YRZ model $\alpha = 1, \Sigma_{\text{corr}}(k) = -\Delta_{\text{pg}}^2 G_0(-k)$, and for the $t$-matrix model $\alpha = 1, \Sigma_{\text{corr}}(k) = \sum_l t(l) G_{\text{BCS}}(l-k)$.

The bare Ward-Takahashi identity is $q_\mu \gamma^\mu(k_+, k_-) = G_0^{-1}(k_+) - G_0^{-1}(k_-)$. Using this, it follows that the Ward-Takahashi identity for the full vertex is [5]

$$\begin{aligned} q_\mu \Gamma^\mu(k_+, k_-) &= G^{-1}(k_+) - G^{-1}(k_-), \\ &= q_\mu \gamma^\mu(k_+, k_-) + \Sigma(k_-) - \Sigma(k_+). \end{aligned} \tag{3}$$

As discussed in the main text, both the strong correlation self-energy $\Sigma_{\text{corr}}$, and the superconducting self-energy $\Sigma_{\text{sc}}$ will give rise to vertex contributions. We therefore write $\Sigma(k) = \Sigma_{\text{corr}}(k) + \Sigma_{\text{sc}}(k)$ and derive the vertex contributions from both self-energies separately. The strong correlation self-energy gives a vertex contribution $\Lambda^\mu(k_+, k_-)$ defined through

$$q_\mu \Lambda^\mu(k_+, k_-) = \Sigma_{\text{corr}}(k_-) - \Sigma_{\text{corr}}(k_+). \tag{4}$$

The general form of this vertex depends on the specific model under consideration. In Sec. (III) we will derive the explicit form of this vertex for three models of interest in the literature. The superconducting vertex is defined through

$$q_\mu \Gamma_{\text{sc}}^\mu(k_+, k_-) = \Sigma_{\text{sc}}(k_-) - \Sigma_{\text{sc}}(k_+). \tag{5}$$

Using these definitions, the full vertex is then

$$\Gamma^\mu(k_+, k_-) = \gamma^\mu(k_+, k_-) + \Lambda^\mu(k_+, k_-) + \Gamma_{\text{sc}}^\mu(k_+, k_-), \tag{6}$$

which can be found from the full Ward-Takahashi identity in Eq. (3).

We now derive the explicit form of $\Gamma_{\text{sc}}^\mu$. The superconducting vertex contributions are most easily found by defining the collective mode vertices $\Pi^\mu(q)$ and $\bar{\Pi}^\mu(q)$ such that $q_\mu \Pi^\mu(q) = 2\Delta_{\text{sc}}$, $q_\mu \bar{\Pi}^\mu(q) = -2\Delta_{\text{sc}}^*$. For now, these will be left as a definition, but these relations, along with the explicit form of $\Pi^\mu$, $\bar{\Pi}^\mu$, will be derived in Sec. (II). Using the superconducting self-energy given in Eq. (2), we then have

$$\Sigma_{\text{sc}}(k_-) - \Sigma_{\text{sc}}(k_+) = -\Delta_{\text{sc}}^* q_\mu \Pi^\mu(q) G_0^\alpha(-k_-) - \Delta_{\text{sc}} q_\mu \bar{\Pi}^\mu(q) G_0^\alpha(-k_+) - |\Delta_{\text{sc}}|^2 \left[G_0^\alpha(-k_+) - G_0^\alpha(-k_-)\right]. \tag{7}$$

The difference of the two partially dressed Green's functions is

$$\begin{aligned} G_0^\alpha(-k_+) - G_0^\alpha(-k_-) &= G_0^\alpha(-k_-) \left[G_0^{-1}(-k_-) - G_0^{-1}(-k_+) + \alpha\left(\Sigma_{\text{corr}}(-k_+) - \Sigma_{\text{corr}}(-k_-)\right)\right] G_0^\alpha(-k_+), \\ &= G_0^\alpha(-k_-) \left[q_\mu \gamma^\mu(-k_-, -k_+) + \alpha q_\mu \Lambda^\mu(-k_-, -k_+)\right] G_0^\alpha(-k_+). \end{aligned} \tag{8}$$

In the second line we have used both the bare Ward-Takahashi identity, as well as the definition of the $\Lambda^\mu$ vertex. Substituting Eq. (7) and Eq. (8) into Eq. (5) then gives the superconducting vertex:

$$\Gamma^\mu_{\rm sc}(k_+, k_-) = -\Delta^*_{\rm sc}\Pi^\mu(q)G_0^\alpha(-k_-) - \Delta_{\rm sc}\bar\Pi^\mu(q)G_0^\alpha(-k_+)$$
$$- |\Delta_{\rm sc}|^2 G_0^\alpha(-k_-)\left[\gamma^\mu(-k_-, -k_+) + \alpha\Lambda^\mu(-k_-, -k_+)\right]G_0^\alpha(-k_+). \tag{9}$$

This then produces the exact gauge invariant full vertex given in Eq. (2.8) of the main text:

$$\Gamma^\mu(k_+, k_-) = \gamma^\mu(k_+, k_-) + \Lambda^\mu(k_+, k_-) - \Delta^*_{\rm sc}\Pi^\mu(q)G_0^\alpha(-k_-) - \Delta_{\rm sc}\bar\Pi^\mu(q)G_0^\alpha(-k_+)$$
$$- |\Delta_{\rm sc}|^2 G_0^\alpha(-k_-)[\gamma^\mu(-k_-, -k_+) + \alpha\Lambda^\mu(-k_-, -k_+)]G_0^\alpha(-k_+). \tag{10}$$

From the above expression it is clear that if $\Sigma_{\rm corr} = 0 \Rightarrow \Lambda^\mu = 0$, then the full vertex reduces to the BCS full vertex [6, 7]. Similarly if $\Delta_{\rm sc} = 0$, then the full vertex reduces to the paired normal state vertex [8, 9]. In order to uniquely determine the full vertex, the collective mode vertices $\Pi^\mu(q)$, $\bar\Pi^\mu(q)$ and the vertex correction $\Lambda^\mu(k_+, k_-)$ must be determined. The Feynman diagrams for the full response function are given in Fig. (1).

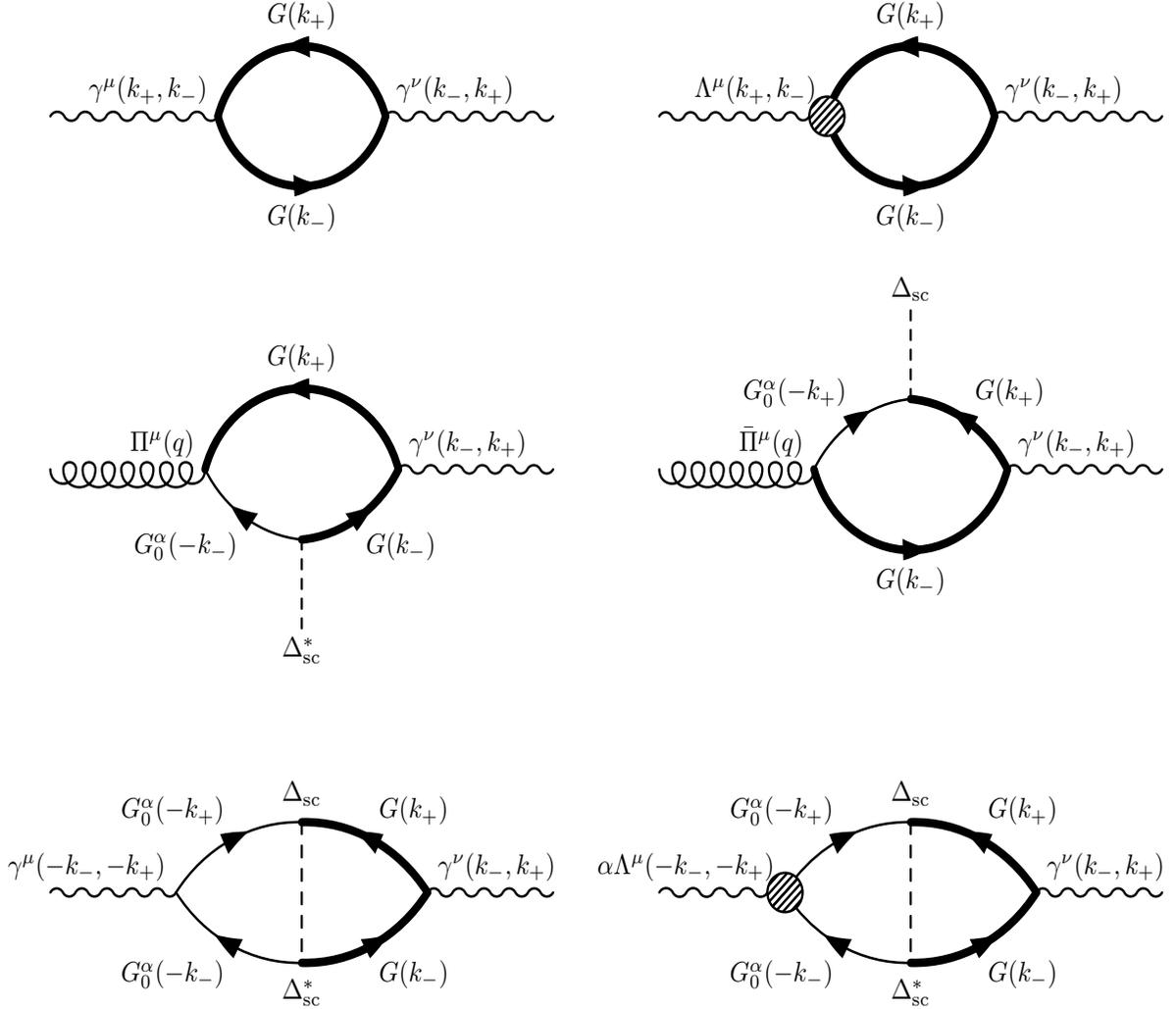

FIG. 1. Feynman diagrams for the two particle response function $P^{\mu\nu}(q) = 2\sum_k G(k_+)\Gamma^\mu(k_+, k_-)G(k_-)\gamma^\nu(k_-, k_+)$ given a self-energy of the form in Eq. (2). The order of appearance of the diagrams from left to right and top to bottom corresponds directly to the order of appearance of terms in Eq. (10). The pseudogap approximation corresponds to $\alpha = 0, \Sigma_{\rm corr}(k) = -\Delta^2_{\rm pg}G_0(-k)$, for the YRZ model $\alpha = 1, \Sigma_{\rm corr}(k) = -\Delta^2_{\rm pg}G_0(-k)$, and for the $t$-matrix model $\alpha = 1, \Sigma_{\rm corr}(k) = \sum_l t(l)G_{\rm BCS}(l-k)$.

## II. COLLECTIVE MODE VERTICES

In this section verify that the collective mode vertices $\Pi^\mu(q)$ and $\bar{\Pi}^\mu(q)$ satisfy the gauge invariant conditions $q_\mu \Pi^\mu(q) = 2\Delta_{\rm sc}$, $q_\mu \bar{\Pi}^\mu(q) = -2\Delta_{\rm sc}^*$, which was assumed in their definitions. These vertices are conveniently expressed as a matrix equation if we define the two-point correlation functions

$$Q_{+-}(q) = 1/g - \sum_k G(k_+) G_0^\alpha(-k_-) \left[1 - \Delta_{\rm sc}^* F_{\rm sc}(k_-)\right],$$

$$Q_{-+}(q) = 1/g - \sum_k G(k_-) G_0^\alpha(-k_+) \left[1 - \Delta_{\rm sc} F_{\rm sc}^*(k_+)\right],$$

$$Q_{++}(q) = \sum_k F_{\rm sc}(k_+) F_{\rm sc}(k_-) = Q_{--}^*(q),$$

$$P_+^\mu(q) = \sum_k \left[\gamma^\mu(k_+,k_-) + \Lambda^\mu(k_+,k_-)\right] G(k_+) F_{\rm sc}(k_-)$$
$$+ \sum_k \left[\gamma^\mu(-k_-,-k_+) + \alpha \Lambda^\mu(-k_-,-k_+)\right] F_{\rm sc}(k_+) G_0^\alpha(-k_-) \left[1 - \Delta_{\rm sc}^* F_{\rm sc}(k_-)\right],$$

$$P_-^\mu(q) = \sum_k \left[\gamma^\mu(k_+,k_-) + \Lambda^\mu(k_+,k_-)\right] F_{\rm sc}^*(k_+) G(k_-)$$
$$+ \sum_k \left[\gamma^\mu(-k_-,-k_+) + \alpha \Lambda^\mu(-k_-,-k_+)\right] G_0^\alpha(-k_+) F_{\rm sc}^*(k_-) \left[1 - \Delta_{\rm sc} F_{\rm sc}^*(k_+)\right]. \tag{11}$$

From Eqs. (2.10-2.11) of the main text, the collective modes can then be written as

$$\begin{pmatrix} \Pi^\mu \\ \bar{\Pi}^\mu \end{pmatrix} = \begin{pmatrix} Q_{+-} & Q_{++} \\ Q_{--} & Q_{-+} \end{pmatrix}^{-1} \begin{pmatrix} P_+^\mu \\ P_-^\mu \end{pmatrix}. \tag{12}$$

We now contract each side of Eq. (12) with $q_\mu$. In order to calculate the right-hand side, we calculate the contraction $q_\mu P_\pm^\mu(q)$:

$$q_\mu P_+^\mu(q) = q_\mu \sum_k \left[\gamma^\mu(k_+,k_-) + \Lambda^\mu(k_+,k_-)\right] G(k_+) F_{\rm sc}(k_-)$$
$$+ q_\mu \sum_k \left[\gamma^\mu(-k_-,-k_+) + \alpha \Lambda^\mu(-k_-,-k_+)\right] F_{\rm sc}(k_+) G_0^\alpha(-k_-) [1 - \Delta_{\rm sc}^* F_{\rm sc}(k_-)]. \tag{13}$$

Explicit calculation shows that both lines have the same value, so that

$$q_\mu P_+^\mu(q) = 2 \left[\Delta_{\rm sc} \left(1/g - \sum_k G(k_+) G_0^\alpha(-k_-)[1 - \Delta_{\rm sc}^* F_{\rm sc}(k_-)]\right) - \Delta_{\rm sc}^* \sum_k F_{\rm sc}(k_-) F_{\rm sc}(k_+)\right],$$
$$= 2 \left(\Delta_{\rm sc} Q_{+-} - \Delta_{\rm sc}^* Q_{++}\right). \tag{14}$$

Similarly, since $\Delta_{\rm sc}^* P_+^\mu(q) = \Delta_{\rm sc} P_-^\mu(-q)$, we also find $q_\mu P_-^\mu(q) = -\left(q_\mu P_+^\mu(q)\right)^*$. The contractions of the collective mode vertices are then

$$\begin{pmatrix} q_\mu \Pi^\mu \\ q_\mu \bar{\Pi}^\mu \end{pmatrix} = \begin{pmatrix} Q_{+-} & Q_{++} \\ Q_{--} & Q_{-+} \end{pmatrix}^{-1} \begin{pmatrix} 2(\Delta_{\rm sc} Q_{+-} - \Delta_{\rm sc}^* Q_{++}) \\ -2(\Delta_{\rm sc}^* Q_{-+} - \Delta_{\rm sc} Q_{--}) \end{pmatrix} = \begin{pmatrix} 2\Delta_{\rm sc} \\ -2\Delta_{\rm sc}^* \end{pmatrix}. \tag{15}$$

This confirms that, for all $q$, we have the desired relations

$$q_\mu \Pi^\mu(q) = 2\Delta_{\rm sc}, \quad q_\mu \bar{\Pi}^\mu(q) = -2\Delta_{\rm sc}^*. \tag{16}$$

Finally, we now show that at $q = 0$ the gap equation is consistent with the poles of the collective mode vertices. These poles are given by the solution of $\det(Q_{ab}) = Q_{+-} Q_{-+} - Q_{++} Q_{--} = 0$, which arises when taking the matrix inverse of Eq. (12). Let $q = 0$, and suppose $\Delta_{\rm sc} = \Delta_{\rm sc}^*$, then the poles occur when $Q_{+-} - Q_{++} = 0$. Using the expressions in Eq. (11), and the definition of $F_{\rm sc}(k)$, this reduces to

$$1 - g \sum_k G_0^\alpha(-k) G(k) = 0, \tag{17}$$

which is the expected gap equation.

In summary, we have obtained the collective mode vertices, and thus obtained the gauge invariant full vertex. The next section determines the form of the vertex $\Lambda^\mu$ for three example cases of $\Sigma_{\rm corr}$.

## III. SPECIFIC EXAMPLES FOR THE $\Lambda^\mu$ VERTEX

### A. Pairing pseudogap approximation

In the pairing pseudogap approximation [1, 2], $\Sigma_{\text{corr}}(k) = -\Delta_{\text{pg}}^2 G_0(-k)$, which implies that

$$q_\mu \Lambda^\mu(k_+, k_-) = \Sigma_{\text{corr}}(k_-) - \Sigma_{\text{corr}}(k_+),$$
$$= \Delta_{\text{pg}}^2 G_0(-k_-) q_\mu \gamma^\mu(-k_-, -k_+) G_0(-k_+). \quad (18)$$

Thus, it follows that

$$\Lambda^\mu(k_+, k_-) = \Delta_{\text{pg}}^2 G_0(-k_-) \gamma^\mu(-k_-, -k_+) G_0(-k_+). \quad (19)$$

### B. YRZ

In the YRZ model [3], $\Sigma_{\text{corr}}(k) = -\Delta_{\text{pg}}^2 G_0(-k)$. Thus, as in the case for the pseudogap approximation, we obtain

$$\Lambda^\mu(k_+, k_-) = \Delta_{\text{pg}}^2 G_0(-k_-) \gamma^\mu(-k_-, -k_+) G_0(-k_+). \quad (20)$$

### C. Particle-only $t$-matrix

In the $t$-matrix model of Ref. [4], $\Sigma_{\text{corr}}(k) = \sum_l t(l) G_{\text{BCS}}(l-k) = \sum_l t(l+k) G_{\text{BCS}}(l)$. Here $G_{\text{BCS}}(k)$ is a BCS Green's function, where we define

$$G_{\text{BCS}}^{-1}(k) = G_0^{-1}(k) - \Sigma_{\text{BCS}}(k),$$
$$F_{\text{BCS}}(k) = \Delta_{\text{mf}} G_0(-k) G_{\text{BCS}}(k). \quad (21)$$

The BCS self-energy is $\Sigma_{\text{BCS}}(k) = -|\Delta_{\text{mf}}|^2 G_0(-k)$, where $\Delta_{\text{mf}}$ is the mean-field BCS gap. The anomalous Green's function satisfies $F_{\text{BCS}}(k) = F_{\text{BCS}}(-k)$.

The inverse $t$-matrix $t^{-1}(l)$ in the formalism of Ref. [4] is given by

$$t^{-1}(l) = \chi_{11}(l) - \chi_{22}^{-1}(l) \chi_{12}(l) \chi_{21}(l), \quad (22)$$

where the susceptibilities are given by

$$\chi_{11}(l) = \frac{1}{g} - \sum_m G_{\text{BCS}}(l+m) G_{\text{BCS}}(-m),$$
$$\chi_{12}(l) = \sum_m F_{\text{BCS}}(l+m) F_{\text{BCS}}^*(-m), \quad (23)$$

where $1/g$ is the standard $s$-wave interaction [4]. Note that $\chi_{22}(l) = \chi_{11}(-l)$, and because $F_{\text{BCS}}(m)$ is even, $\chi_{12}(l) = \chi_{21}^*(l) = \chi_{21}(l)$.

The vertex correction $\Lambda^\mu$ for this model is defined by Eq. (4). We now proceed to evaluate the right hand side of Eq. (4). From the definition of $\Sigma_{\text{corr}}$, it follows that

$$\Sigma_{\text{corr}}(k_-) - \Sigma_{\text{corr}}(k_+) = 2 \sum_l G_{\text{BCS}}(l) t(l+k_+) t(l+k_-) \left( t^{-1}(l+k_+) - t^{-1}(l+k_-) \right)$$
$$- \sum_l t(l) G_{\text{BCS}}(l-k_+) G_{\text{BCS}}(l-k_-) \left( G_{\text{BCS}}^{-1}(l-k_+) - G_{\text{BCS}}^{-1}(l-k_-) \right). \quad (24)$$

The BCS Green's function also obeys a Ward-Takahashi identity, which defines the BCS full vertex $\Gamma_{\text{BCS}}^\mu$:

$$q_\mu \Gamma_{\text{BCS}}^\mu(k_+, k_-) = G_{\text{BCS}}^{-1}(k_+) - G_{\text{BCS}}^{-1}(k_-). \quad (25)$$

Equivalently, $\Gamma_{\text{BCS}}^\mu$ is given by Eq. (10) with $\Sigma_{\text{corr}} = 0 = \Lambda^\mu$. Thus, we now have

$$\Sigma_{\text{corr}}(k_-) - \Sigma_{\text{corr}}(k_+) = 2 \sum_l G_{\text{BCS}}(l) t(l+k_+) t(l+k_-) \left( t^{-1}(l+k_+) - t^{-1}(l+k_-) \right)$$
$$+ \sum_l t(l) G_{\text{BCS}}(l-k_-) q_\mu \Gamma_{\text{BCS}}^\mu(l-k_-, l-k_+) G_{\text{BCS}}(l-k_+). \quad (26)$$

From the $t$-matrix definition in Eq. (22), the difference of the two inverse $t$-matrices is

$$t^{-1}(l+k_+) - t^{-1}(l+k_-) = \chi_{11}(l+k_+) - \chi_{11}(l+k_-)$$
$$+ \chi_{22}^{-1}(l+k_-)\chi_{12}(l+k_-)\chi_{21}(l+k_-) - \chi_{22}^{-1}(l+k_+)\chi_{12}(l+k_+)\chi_{21}(l+k_+). \qquad (27)$$

For the first line of this expression we can use the Ward-Takahashi identity in Eq. (25) to obtain

$$\chi_{11}(l+k_+) - \chi_{11}(l+k_-) = \sum_m G_{\text{BCS}}(-m)G_{\text{BCS}}(l+m+k_+)q_\mu \Gamma_{\text{BCS}}^\mu(l+m+k_+, l+m+k_-)G_{\text{BCS}}(l+m+k_-). \qquad (28)$$

It remains to compute the difference term in the second line of Eq. (27). To do this, first note that

$$\chi_{22}^{-1}(l+k_-)\chi_{12}(l+k_-)\chi_{21}(l+k_-) - \chi_{22}^{-1}(l+k_+)\chi_{12}(l+k_+)\chi_{21}(l+k_+)$$
$$= \Big\{ [\chi_{22}(l+k_+) - \chi_{22}(l+k_-)]\chi_{12}(l+k_-)\chi_{21}(l+k_-)$$
$$+ [\chi_{12}(l+k_-) - \chi_{12}(l+k_+)]\chi_{22}(l+k_-)\chi_{21}(l+k_-)$$
$$+ [\chi_{21}(l+k_-) - \chi_{21}(l+k_+)]\chi_{22}(l+k_-)\chi_{12}(l+k_+) \Big\} (\chi_{22}(l+k_-)\chi_{22}(l+k_+))^{-1}. \qquad (29)$$

This form simplifies the problem to computing the vertex insertions into both $\chi_{12}, \chi_{21}$, and $\chi_{22}$ individually, and then summing the result. Since $\chi_{22}(k) = \chi_{11}(-k)$, we can use the result in Eq. (28) to obtain

$$\chi_{22}(l+k_+) - \chi_{22}(l+k_-) = -\sum_m G_{\text{BCS}}(m)G_{\text{BCS}}(-l-m-k_-)q_\mu \Gamma_{\text{BCS}}^\mu(-l-m-k_-, -l-m-k_+)G_{\text{BCS}}(-l-m-k_+). \qquad (30)$$

We now study the $\chi_{12}$ difference term in the third line of Eq. (29). This difference amounts to performing all possible vertex insertions into $\chi_{12}$. If we write $\chi_{12}(k) = |\Delta_{\text{mf}}|^2 \sum_m G_0(-m-k)G_{\text{BCS}}(m+k)G_0(m)G_{\text{BCS}}(-m)$, then it is clear that there are six possible positions for vertex insertions; two full vertices can be inserted into the full Green's functions, two bare vertices can be inserted into the bare Green's functions, and two collective mode vertices can be inserted into the fluctuating gap $\Delta_{\text{mf}}$ or $\Delta_{\text{mf}}^*$. Performing all these vertex insertions then gives the following result:

$$2(\chi_{12}(l+k_+) - \chi_{12}(l+k_-))$$
$$= q_\mu \bar{\Pi}_{\text{mf}}^\mu(q) \sum_m G_0(m+q)G_{\text{BCS}}(-m)F(m+l+k_+) + q_\mu \Pi_{\text{mf}}^\mu(q) \sum_m G_0(-m-l-k_-)G_{\text{BCS}}(m+l+k_+)F^*(m)$$
$$+ \sum_m F(m+l+k_+)[G_0(m+q)q_\mu\gamma^\mu(m+q,m)F^*(m) - G_{\text{BCS}}(-m)q_\mu\Gamma^\mu(-m,-m-q)F^*(m+q)]$$
$$+ \sum_m F^*(m)[G_0(-m-l-k_-)q_\mu\gamma^\mu(-m-l-k_-,-m-l-k_+)F(m+l+k_+)$$
$$- G_{\text{BCS}}(m+l+k_+)q_\mu\Gamma^\mu(m+l+k_+, m+l+k_-)F(m+l+k_-)]. \qquad (31)$$

Here we have introduced the collective mode vertices $\Pi_{\text{mf}}^\mu(q), \bar{\Pi}_{\text{mf}}^\mu(q)$, which satisfy $q_\mu \Pi_{\text{mf}}^\mu(q) = 2\Delta_{\text{mf}}, q_\mu \bar{\Pi}_{\text{mf}}^\mu(q) = -2\Delta_{\text{mf}}^*$. This is the mean-field BCS version of the collective mode vertices discussed in Sec. (II). Since $\chi_{12} = \chi_{21}$, the same result derived above holds for $\chi_{21}$. We can now combine all the previous results from this subsection and define the following vertices

$$v_{11}^\mu(l+k_+, l+k_-) = \sum_m G_{\text{BCS}}(-m)G_{\text{BCS}}(l+m+k_+)\Gamma_{\text{BCS}}^\mu(l+m+k_+, l+m+k_-)G_{\text{BCS}}(l+m+k_-)$$
$$+ \sum_m G_{\text{BCS}}(l+m+k_+)G_{\text{BCS}}(-m)\Gamma_{\text{BCS}}^\mu(-m,-m-q)G_{\text{BCS}}(-m-q). \qquad (32)$$

$$v_{22}^\mu(l+k_+, l+k_-) = -\frac{\chi_{12}(l+k_-)\chi_{21}(l+k_-)}{\chi_{22}(l+k_-)\chi_{22}(l+k_+)}$$
$$\times \Bigg[\sum_m G_{\text{BCS}}(m)G_{\text{BCS}}(-l-m-k_-)\Gamma_{\text{BCS}}^\mu(-l-m-k_-, -l-m-k_+)G_{\text{BCS}}(-l-m-k_+)$$
$$+ \sum_m G_{\text{BCS}}(-l-m-k_-)G_{\text{BCS}}(m)\Gamma_{\text{BCS}}^\mu(m, m-q)G_{\text{BCS}}(m-q)\Bigg]. \qquad (33)$$

$$v_{12}^{\mu}(l+k_+, l+k_-) = -\frac{\chi_{21}(l+k_-)}{\chi_{22}(l+k_+)}\left\{\bar{\Pi}_{\text{mf}}^{\mu}(q)\sum_m G_0(m+q)G_{\text{BCS}}(-m)F(m+l+k_+)\right.$$

$$+ \Pi_{\text{mf}}^{\mu}(q)\sum_m G_0(-m-l-k_-)G_{\text{BCS}}(m+l+k_+)F^*(m)$$

$$+ \sum_m F(m+l+k_+)\left[G_0(m+q)\gamma^{\mu}(m+q,m)F^*(m) - G_{\text{BCS}}(-m)\Gamma^{\mu}(-m,-m-q)F^*(m+q)\right]$$

$$+ \sum_m F^*(m)[G_0(-m-l-k_-)\gamma^{\mu}(-m-l-k_-,-m-l-k_+)F(m+l+k_+)$$

$$\left.- G_{\text{BCS}}(m+l+k_+)\Gamma^{\mu}(m+l+k_+,m+l+k_-)F(m+l+k_-)]\right\}. \tag{34}$$

$$v_{21}^{\mu}(l+k_+, l+k_-) = \frac{\chi_{12}(l+k_+)}{\chi_{21}(l+k_-)}v_{12}^{\mu}(l+k_+, l+k_-). \tag{35}$$

Using the definitions of these vertices, along with Eq. (24), finally gives the vertex $\Lambda^{\mu}(k_+, k_-)$ for the $t$-matrix model of Ref. [4]

$$\Lambda^{\mu}(k_+, k_-) = \sum_l t(l)G_{\text{BCS}}(l-k_-)\Gamma_{\text{BCS}}^{\mu}(l-k_-, l-k_+)G_{\text{BCS}}(l-k_+)$$

$$+ \sum_l G(l)t(l+k_+)\left[v_{11}^{\mu}(l+k_+, l+k_-) + v_{12}^{\mu}(l+k_+, l+k_-)\right.$$

$$\left.+ v_{21}^{\mu}(l+k_+, l+k_-) + v_{22}^{\mu}(l+k_+, l+k_-)\right]t(l+k_-). \tag{36}$$

It can be shown that this vertex does indeed satisfy $q_{\mu}\Lambda^{\mu}(k_+, k_-) = \Sigma_{\text{corr}}(k_-) - \Sigma_{\text{corr}}(k_+)$. Diagrammatically, the first line in this expression is a Maki-Thompson (MT) diagram. The first term in the parentheses of the second line represents two identical Aslamazov-Larkin (AL) diagrams [7]. Similarly the fourth term in parentheses is similar to two identitcal Aslamazov-Larkin diagrams. The second and third terms in parentheses are additional diagrams which must be retained in order to satisfy the gauge invariant condition $q_{\mu}\Lambda^{\mu}(k_+, k_-) = \Sigma_{\text{corr}}(k_-) - \Sigma_{\text{corr}}(k_+)$. In the normal state $(T > T_c)$ $v_{12} = v_{21} = v_{22} = 0$ and the above vertex is then the familiar MT + 2AL diagrams.

## IV. $f$-SUM RULE AND LONGITUDINAL SUM RULE

In this section we show that, given bare and full vertices that satisfy the Ward-Takahashi identity, the density-density and current-current response functions satisfy the $f$ and longitudinal sum rules, respectively.

The exact response function is constructed from a two point correlation function containing one full vertex, $\Gamma^{\mu}(k_+, k_-) = (\Gamma^0(k_+, k_-), \boldsymbol{\Gamma}(k_+, k_-))$, and one bare vertex $\gamma^{\nu}(k_+, k_-) = (\gamma^0(k_+, k_-), \boldsymbol{\gamma}(k_+, k_-))$:

$$P^{\mu\nu}(q) = 2\sum_k G(k_+)\Gamma^{\mu}(k_+, k_-)G(k_-)\gamma^{\nu}(k_-, k_+). \tag{37}$$

To show consistency with sum rules, we use the Ward-Takahashi identity (as in Eq. (3))

$$q_{\mu}\Gamma^{\mu}(k_+, k_-) = G^{-1}(k_+) - G^{-1}(k_-). \tag{38}$$

Contracting the response function with $q_{\mu}$, and using the Ward-Takahashi identity, we then have

$$q_{\mu}P^{\mu\nu}(q) = 2\sum_k G(k_+)[G^{-1}(k_+) - G^{-1}(k_-)]G(k_-)\gamma^{\nu}(k_-, k_+),$$

$$= 2\sum_k G(k)[\gamma^{\nu}(k, k+q) - \gamma^{\nu}(k-q, k)]. \tag{39}$$

The $\nu = 0$ component of the bare vertex is equal to one, so that

$$q_{\mu}P^{\mu 0}(q) = 0 \tag{40}$$

On the other hand, the spatial components $\nu = j \in \{1, 2, 3\}$ are

$$q_\mu P^{\mu j}(q) = 2\sum_k G(k)[\boldsymbol{\gamma}(k, k+q) - \boldsymbol{\gamma}(k-q, k)],$$
$$= \frac{n}{m}\mathbf{q}. \tag{41}$$

Here we used the fact that $\boldsymbol{\gamma}(k, k+q) - \boldsymbol{\gamma}(k-q, k) = \mathbf{q}/m$ is independent of $\mathbf{k}$, and $2\sum_k G(k) = 2\sum_\mathbf{k} n_\mathbf{k} = n$.

In terms of components, and real frequencies, these equations become

$$\omega P^{00}(\omega, \mathbf{q}) - \mathbf{q} \cdot \mathbf{P}^{i0}(\omega, \mathbf{q}) = 0, \tag{42}$$
$$\omega \mathbf{P}^{0j}(\omega, \mathbf{q}) - \mathbf{q} \cdot \overset{\leftrightarrow}{P}{}^{ij}(\omega, \mathbf{q}) = \frac{n}{m}\mathbf{q}. \tag{43}$$

Setting $\omega = 0$ and then operating with $-\mathbf{q}$ (on the right) in Eq. (43) gives

$$\mathbf{q} \cdot \overset{\leftrightarrow}{P}{}^{ij}(0, \mathbf{q}) \cdot \mathbf{q} = -\frac{n}{m}\mathbf{q} \cdot \mathbf{q}. \tag{44}$$

Now use the identity Im $P^{i0}(\omega, \mathbf{q}) = -$Im $P^{0i}(-\omega, -\mathbf{q})$ and Eq. (42), Eq. (43) to solve for Im $P^{00}$ in terms of Im $\overset{\leftrightarrow}{P}{}^{ij}$. Applying the Kramers-Kronig relations and Eq. (44) then gives

$$\int \frac{d\omega}{\pi}(-\omega\text{Im } P^{00}(\omega, \mathbf{q})) = \int \frac{d\omega}{\pi}\left(-\frac{\mathbf{q} \cdot \text{Im } \overset{\leftrightarrow}{P}{}^{ij}(\omega, \mathbf{q}) \cdot \mathbf{q}}{\omega}\right)$$
$$= -\mathbf{q} \cdot \text{Re } \overset{\leftrightarrow}{P}{}^{ij}(0, \mathbf{q}) \cdot \mathbf{q}$$
$$= \frac{n}{m}\mathbf{q} \cdot \mathbf{q}. \tag{45}$$

The density-density and current-current response functions are respectively defined by $\chi_{\rho\rho}(q) \equiv P^{00}(q)$, $\overset{\leftrightarrow}{\chi}_{JJ}(q) \equiv P^{ij}(q), i, j \in \{1, 2, 3\}$. The $f$-sum rule is then

$$\int \frac{d\omega}{\pi}(-\omega\text{Im } \chi_{\rho\rho}(\omega, \mathbf{q})) = \frac{n}{m}\mathbf{q} \cdot \mathbf{q}. \tag{46}$$

Similarly the longitudinal sum rule is

$$\int \frac{d\omega}{\pi}\left(-\frac{\mathbf{q} \cdot \text{Im } \overset{\leftrightarrow}{\chi}_{JJ}(\omega, \mathbf{q}) \cdot \mathbf{q}}{\omega}\right) = \frac{n}{m}\mathbf{q} \cdot \mathbf{q}. \tag{47}$$

Therefore, provided the full vertex satisfies the Ward-Takahashi identity, both sum rules will hold exactly for all $\mathbf{q}$ in this continuum limit.


\end{document}